\documentclass[prx,aps,twocolumn]{revtex4}
\usepackage{graphicx}
\usepackage{epsfig}
\usepackage{enumerate}
\usepackage{amssymb}
\usepackage{amsmath}
\usepackage{amsfonts}
\usepackage{dcolumn}
\usepackage{bm}
\usepackage{framed}
\usepackage{braket}
\usepackage{color}

\begin{document}
\title{Theory of graphene saturable absorption}
\author{A. Marini$^{1}$}
\email{andrea.marini@icfo.es}
\author{J. D. Cox$^{1}$}
\author{F.~J.~Garc\'{\i}a~de~Abajo$^{1,2}$}
\affiliation{$^1$ICFO-Institut de Ciencies Fotoniques, The Barcelona Institute of Science and Technology, 08860 Castelldefels (Barcelona), Spain}
\affiliation{$^2$ICREA-Instituci\'o Catalana de Recerca i Estudis Avan\c{c}ats, Passeig Llu\'is Companys 23, 08010 Barcelona, Spain}
\begin{abstract}
Saturable absorption is a non-perturbative nonlinear optical phenomenon that plays a pivotal role in the generation of ultrafast light pulses. Here we show that this effect emerges in graphene at unprecedentedly low light intensities, thus opening avenues to new nonlinear physics and applications in optical technology. Specifically, we theoretically investigate saturable absorption in extended graphene by developing a non-perturbative single-particle approach, describing conduction-electron dynamics in the atomically-thin material using the two-dimensional Dirac equation for massless Dirac fermions, which is recast in the form of generalized Bloch equations. By solving the electron dynamics non-perturbatively, we account for both interband and intraband contributions to the intensity-dependent saturated conductivity and conclude that the former dominates regardless of the intrinsic doping state of the material.  The results are in excellent agreement with atomistic quantum-mechanical simulations including higher-band effects. Additionally, we find that the modulation depth of saturable absorption in graphene can be electrically manipulated through an externally applied gate voltage. Our results are relevant for the development of graphene-based optoelectronic devices, as well as for applications in mode-locking and random lasers.
\end{abstract}
\maketitle

\section{Introduction}

Saturable absorption (SA) is an extreme nonlinear phenomenon that consists in the quenching of optical absorption under high-intensity illumination. This effect, which is an inherent property of photonic materials, constitutes a key element for passive mode-locking (PML) in laser cavities \cite{Ippen1994,Paschotta2001}, where continuous waves are broken into a train of ultrashort optical pulses. Most materials undergo saturable absorption at very high optical intensities, in close proximity to their optical damage threshold. Currently, state-of-the-art semiconductor-based SA mirrors are routinely employed for PML lasers \cite{Steinmeyer1999,Xiang2002,Keller2003}. However, these mirrors operate in a narrow spectral range, are poorly tunable, and require advanced fabrication techniques. 

Recently, carbon nanomaterials have emerged as an attractive, viable, and cost-effective alternative for the development of next-generation PML lasers. For example, carbon nanotubes (CNTs) undergo SA at rather modest light intensities, while their operation wavelength (determined by the energy band gap) can be manipulated by tuning their diameter \cite{Rozhin2006,Scardaci2008,Wang2008,Sun2009,Hasan2009}. Broadband operation has been demonstrated by using an ensemble of CNTs with a wide distribution in diameter, at the expense of higher linear loss from off-resonance tubes \cite{Wang2008}. Graphene overcomes this limitation thanks to its peculiar conical band structure, which gives rise to broadband resonant SA at remarkably low light intensity \cite{Hasan2009,Bao2009,Tan2010,Zhangg2010,Xing2010,Martinez2011,Baoo2011,Winzer2012,Baek2013} that can further be tuned by means of an externally applied gate voltage \cite{Giovannetti2008}. Graphene-based SA components have been used to achieve PML ultrafast laser operation \cite{Sun2010,Popa2010}, broadband tunability \cite{Sunn2010}, and quality-factor switching \cite{Popa2011}. Graphene multilayers have also been employed to generate large energy pulses \cite{Zhang2009} and to achieve PML in fiber lasers with normal dispersion \cite{Zhang2010}. In addition, recent theoretical investigations predict single-mode operation of random lasers by embedding graphene flakes in a gain medium \cite{Marini2016}.

Despite the rising interest in SA of graphene for developing the above mentioned applications, a transparent and accurate theory of graphene SA is still missing. This is most likely due to the fact that SA is inherently a non-perturbative phenomenon, which cannot be accurately described by standard perturbative approaches used to calculate nonlinear graphene conductivities \cite{Peres2014,Mikhailov2014,Mikhailov2016}. Furthermore, theoretical approaches relying on the Boltzmann transport equation (accounting only for intraband electron dynamics) \cite{Mikhailov2007,Mikhailov2008,Mikhailovv2008} are inadequate to simulate SA of graphene, which arises due to the interplay of intraband and interband dynamics. In principle, SA can be modeled through time-domain integration of the single-particle density matrix equations in graphene nanoislands \cite{Cox2014}. However, the computational demand of this approach becomes unaffordable when modeling nanostructures with dimensions $>20\,$nm. Then, in this regime, the single-particle massless Dirac fermion (MDF) picture \cite{Novoselov2005} appears to be an appropriate theoretical environment for describing graphene SA.

In this article, we calculate intraband and interband contributions to SA of extended graphene by non-perturbatively solving the single-particle Dirac equation for MDFs in the presence of an external electromagnetic field. We further investigate the dependence of the intensity-saturated graphene conductivity on doping, temperature, and optical frequency. Interestingly, we find a remarkably low intensity threshold for SA, which is consistent with available experimental reports \cite{Hasan2009,Bao2009,Tan2010,Zhangg2010,Xing2010,Martinez2011,Baoo2011,Winzer2012,Baek2013}. Our calculations indicate a strong quenching of absorption depth produced by electrical doping (which can be controlled through gating), as well as a weak dependence on electron temperature. The SA behavior predicted in the MDF picture for extended graphene is found to be in excellent agreement with atomistic time-domain simulations of graphene nanoribbons based on the tight-binding/single-particle density matrix formalism \cite{Cox2014}. We understand that the present findings are highly relevant for the future development of graphene-based PML fibre lasers and single-mode random lasers. 

\section{Theoretical model}
\label{theoreticalmodel}

\begin{figure}[t]
\centering
\begin{center}
\includegraphics[width=0.5\textwidth]{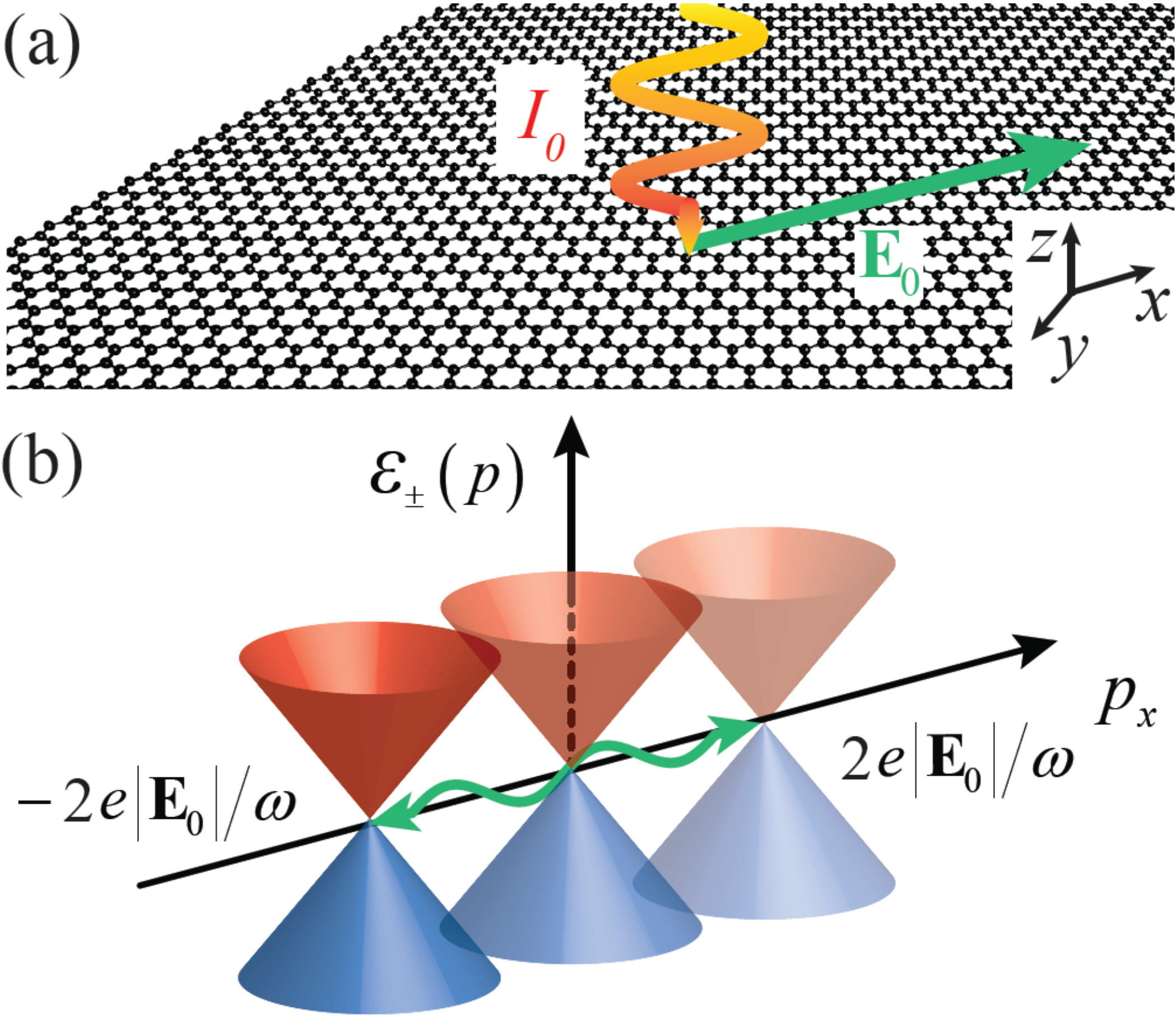}
\caption{(Color online) {\bf (a)} Schematic representation of a light wave (angular frequency $\omega$, amplitude ${\bf E}_0$, and intensity $I_0$) normally impinging on a self-standing graphene sheet. {\bf (b)} Illustration of the temporal displacement of the Dirac cone along the $p_{x}$ direction due to the oscillatory optical field. The central cone represents the unperturbed conical dispersion of graphene ($\varepsilon_{\pm}$), while left and right cones indicate the maximum achievable displacement.}
\label{Fig1}
\end{center}
\end{figure}

We consider a monochromatic optical field ${\bf E}(t) = {\bf E}_0 {\rm e}^{-i\omega t}+{\rm c.c.}$ of angular frequency $\omega$ and complex amplitude ${\bf E}_0$ impinging at normal incidence on a self-standing extended graphene sheet [see Fig.\ \ref{Fig1}(a)]. At visible and lower frequencies, electrons in this material behave as MDFs, with their temporal evolution governed by the single-particle Dirac equation \cite{CastroNetoRMP2009}
\begin{equation}
i\hbar\partial_t \psi_{\bf p}(t) = v_{\rm F} \mbox{\boldmath${\pi}$} \cdot \mbox{\boldmath${\sigma}$} \psi_{\bf p}(t) , \label{DiracEq} 
\end{equation}
where ${\bf p}$ is the electron momentum, $\partial_t$ is the time derivative, $v_{\rm F} \simeq c / 300$ is the Fermi velocity, $c$ is the speed of light in vacuum, $\hbar$ is the reduced Planck constant, $\mbox{\boldmath${\sigma}$} = (\sigma_x, ~ \sigma_y)$ is the two-dimensional (2D) Pauli-matrix vector, and $\psi_{\bf p}(t)$ is the ${\bf p}$- and time-dependent two-component spinor accounting for the quantum states in the upper and lower Dirac-cone bands. Here, we introduce an electron quasi-momentum $\mbox{\boldmath${\pi}$}$ that coincides with the unperturbed momentum (i.e., $\mbox{\boldmath${\pi}$} = {\bf p}$) in the absence of external illumination. In this case, Eq.\ (\ref{DiracEq}) admits spinor eigenvectors
\begin{align}
\psi_{{\bf p},0}^{\pm}(t) =  \frac{1}{\sqrt{2}} \left(\begin{array}{c} {\rm e}^{-i\phi/2}  \\ \pm {\rm e}^{ i\phi/2} \end{array}\right) {\rm e}^{-i\varepsilon_{\pm} t},  \nonumber
\end{align}
where $\hbar\varepsilon_{\pm} = \pm v_{\rm F} p$ is the unperturbed conical dispersion of upper ($+$) and lower ($-$) energy bands [see Fig.\ \ref{Fig1}(b)], while $\phi$ identifies the momentum direction, such that ${\bf p} = p (\cos\phi,\sin\phi)$, and a spatial dependence ${\rm e}^{i {\bf p}\cdot{\bf r}/\hbar}/\sqrt{\cal A}$ (normalized to the sheet area ${\cal A}$) is understood in the spinor.

In the presence of the impinging optical field, we use the customary minimal electron-light coupling prescription to write the electron quasi-momentum as $\mbox{\boldmath${\pi}$}(t) = {\bf p} + e {\bf A}(t)$, where $-e$ is the electron charge and ${\bf A}(t) = -\int {\bf E}(t)dt$ is the potential vector in the Coulomb gauge ($\nabla \cdot {\bf A} = 0$). Without loss of generality, we assume the external light to be linearly polarized along the in-plane $x$ direction, so that ${\bf E}(t) =E_0 {\rm e}^{-i\omega t} {\bf \hat{x}}+{\rm c.c.}$ and ${\bf A}(t) = (E_0/i\omega) {\rm e}^{-i\omega t}{\bf \hat{x}}+{\rm c.c.}$. Therefore, the incident field induces an oscillatory shift of the unperturbed bands $\varepsilon_{\pm}$ along $p_x$ around ${\bf p} = 0$ [see Fig.\ \ref{Fig1}(b)]. Following the non-perturbative approach developed by Ishikawa \cite{Ishikawa2010,Ishikawa2013}, which we review in this section for the sake of completeness, we write the time-dependent spinor as a linear combination of the instantaneous upper- and lower-cone states
\begin{equation}
\psi_{\bf p}(t) = c_{\bf p}^{+}(t) \psi_{\bf p}^{+}(t) + c_{\bf p}^{-}(t) \psi_{\bf p}^{-}(t) \label{AnsatzEq},  
\end{equation}
where
\begin{align}
\psi_{\bf p}^{\pm}(t) =  \frac{1}{\sqrt{2}} \left(\begin{array}{c} {\rm e}^{-i\theta_{\bf p}(t)/2}  \\ \pm {\rm e}^{ i\theta_{\bf p}(t)/2} \end{array}\right) {\rm e}^{\mp i \Omega_{\bf p}(t)},  
\end{align}
$\Omega_{\bf p}(t) = (v_{\rm F}/\hbar) \int  \left| {\bf p} + e {\bf A}(t) \right| dt$ is a global dynamical phase, and $\theta_{\bf p}(t) = {\rm atan} \{p_y/[p_x + e A(t)]\}$ is the time-dependent direction angle of the electron quasi-momentum $\mbox{\boldmath${\pi}$}$. We now insert the Ansatz provided by Eq.\ (\ref{AnsatzEq}) into Eq.\ (\ref{DiracEq}) and define the interband coherence $\rho_{\bf p} = c_{\bf p}^{+} c_{\bf p}^{-*}$ and the population difference $n_{\bf p} = |c_{\bf p}^{+}|^2 - |c_{\bf p}^{-}|^2$. Without making any approximations, we can then rewrite the two-dimensional Dirac equation for MDFs in the form of generalized Bloch equations (GBEs) \cite{Ishikawa2010,Ishikawa2013},
\begin{subequations}
\begin{align}
\dot{\rho}_{\bf p}(t) & =   - \frac{i}{2} \dot{\theta}_{\bf p}(t)    n_{\bf p}(t) {\rm e}^{ 2i\Omega_{\bf p}(t)} , \label{GBEq1}\\
\dot{n}_{\bf p}(t)    & =   2 \dot{\theta}_{\bf p}(t) {\rm Im} \left\{ \rho_{\bf p}(t) {\rm e}^{-2i\Omega_{\bf p}(t)} \right\}. \label{GBEq2}
\end{align}
\label{GBEqs}
\end{subequations}
The carrier current is thus obtained by subtracting the valence band current (assuming the valence band is fully populated) from the total electron current, since it depends only on the carrier occupation, resulting in \cite{Ishikawa2010,Ishikawa2013}
\begin{align}
{\bf j}_{\bf p}(t) & =  \psi_{\bf p}^{\dagger}(t)\mbox{\boldmath${\sigma}$}\psi_{\bf p}(t) - \psi_{\bf p}^{-\dagger}(t)\mbox{\boldmath${\sigma}$}\psi_{\bf p}^{-}(t) \\
& =  \left[ (n_{\bf p}+1)\cos \theta_{\bf p} - 2 \sin \theta_{\bf p} {\rm Im} \left\{ \rho_{\bf p} {\rm e}^{-2i\Omega_{\bf p}} \right\} \right] {\bf \hat{x}} +  \nonumber \\
& +  \left[ (n_{\bf p}+1)\sin \theta_{\bf p} + 2 \cos \theta_{\bf p} {\rm Im} \left\{ \rho_{\bf p} {\rm e}^{-2i\Omega_{\bf p}} \right\} \right] {\bf \hat{y}}.   \nonumber
\end{align}
We emphasize that this expression is obtained non-perturbatively, without any approximation beyond the MDF picture. The $(n_{\bf p}+1)$ terms account for the intraband current, while the remaining terms, which depend on the coherence $\rho_{\bf p}$, arise from interband dynamics \cite{Ishikawa2010,Ishikawa2013}. The macroscopic induced surface current ${\bf J}(t)$ is finally obtained by integrating over all electron momenta,
\begin{equation}
{\bf J}(t) = - \frac{g_{\rm s}g_{\rm v}ev_{\rm F}}{(2\pi\hbar)^2} \int {\bf j}_{\bf p}(t) d^2{\bf p}, \label{CurDensEq}
\end{equation}
where $g_{\rm s}=g_{\rm v} =2$ account for spin and valley degeneracies, and the integral extends over the entire 2D ${\bf p}$ plane. We remark that, although the carrier current ${\bf j}_{\bf p}$ has both ${\bf \hat{x}}$ and ${\bf \hat{y}}$ components, the integrated macroscopic current remains polarized along ${\bf \hat{x}}$ and thus parallel to the external field. This theoretical framework describes the optical response of graphene in a single-electron picture, but it still neglects inelastic electron transitions that are produced for example by impurity scattering and coupling to phonons. In the following sections, these interactions are introduced phenomenologically through an effective electron lifetime. Additionally, we argue that the response of graphene, quantified as a $2.3\%$ absorption at low intensities, is sufficiently weak as to ignore Coulomb self-interaction among induced charges \cite{NBG08,MSW08}.

\section{Intraband saturable absorption}
\label{intrabandsat}

We focus first on the intraband contribution to saturable absorption, i.e., neglecting interband dynamics. This contribution is dominant in doped graphene for photon energies smaller than twice the Fermi level (i.e., $\hbar \omega < 2 E_{\rm F}$), for which interband transitions are prohibited by the Pauli exclusion principle. Consequently, in this regime, one need not solve explicitly the GBEs, as it is sufficient to set $\rho_{\bf p} = 0$ and $n_{\bf p} = {\cal N}(p) = {\cal F}(p) - {\cal F}(-p)$, where ${\cal F}(p) = 1/\{ 1 + {\rm exp}[(v_{\rm F}p - \mu)/k_{\rm B}T]\}$ is the Fermi-Dirac occupation number, $\mu$ is the chemical potential, $k_{\rm B}$ is the Boltzmann constant, and $T$ is the electron temperature. The resulting macroscopic current then reduces to
\begin{equation}
{\bf J}_{\rm intra}(t) = - \frac{ev_{\rm F}}{\pi^2\hbar^2} {\bf \hat{x}} \int [{\cal N}(p)+1] \cos \theta_{\bf p}(t) d^2{\bf p} \label{IntraIntegralEq}.
\end{equation}
Now, in order to account for inelastic electron collisions, we modify heuristically the definition of the direction angle $\theta_{\bf p}(t)$, assuming that the intraband electron quasi-momentum satisfies the differential equation
\begin{equation}
\dot{\mbox{\boldmath${\pi}$}} + \tau^{-1}(\mbox{\boldmath${\pi}$} - {\bf p}) = - e {\bf E}(t),
\label{pitauintra}
\end{equation}
where $\tau$ is the characteristic inelastic collision time. We assume a value $\tau=22\,$fs throughout this work, which is consistent with recent experiments \cite{Johannsen2013,Gierz2013,Brida2013}. We find Eq.\ (\ref{pitauintra}) to admit the straightforward analytical solution $\mbox{\boldmath${\pi}$}(t) = {\bf p} + e {\bf a}(t)$, where 
\begin{align}
{\bf a}(t) & = - {\rm e}^{-t/\tau} \int_{-\infty}^{t}  {\bf E}(t') {\rm e}^{t'/\tau} dt' \nonumber \\
           & = \frac{E_0 {\rm e}^{-i\omega t}}{i\omega-1/\tau} {\bf \hat{x}}+{\rm c.c.}.
\end{align} 

\begin{figure}[t]
\centering
\begin{center}
\includegraphics[width=0.5\textwidth]{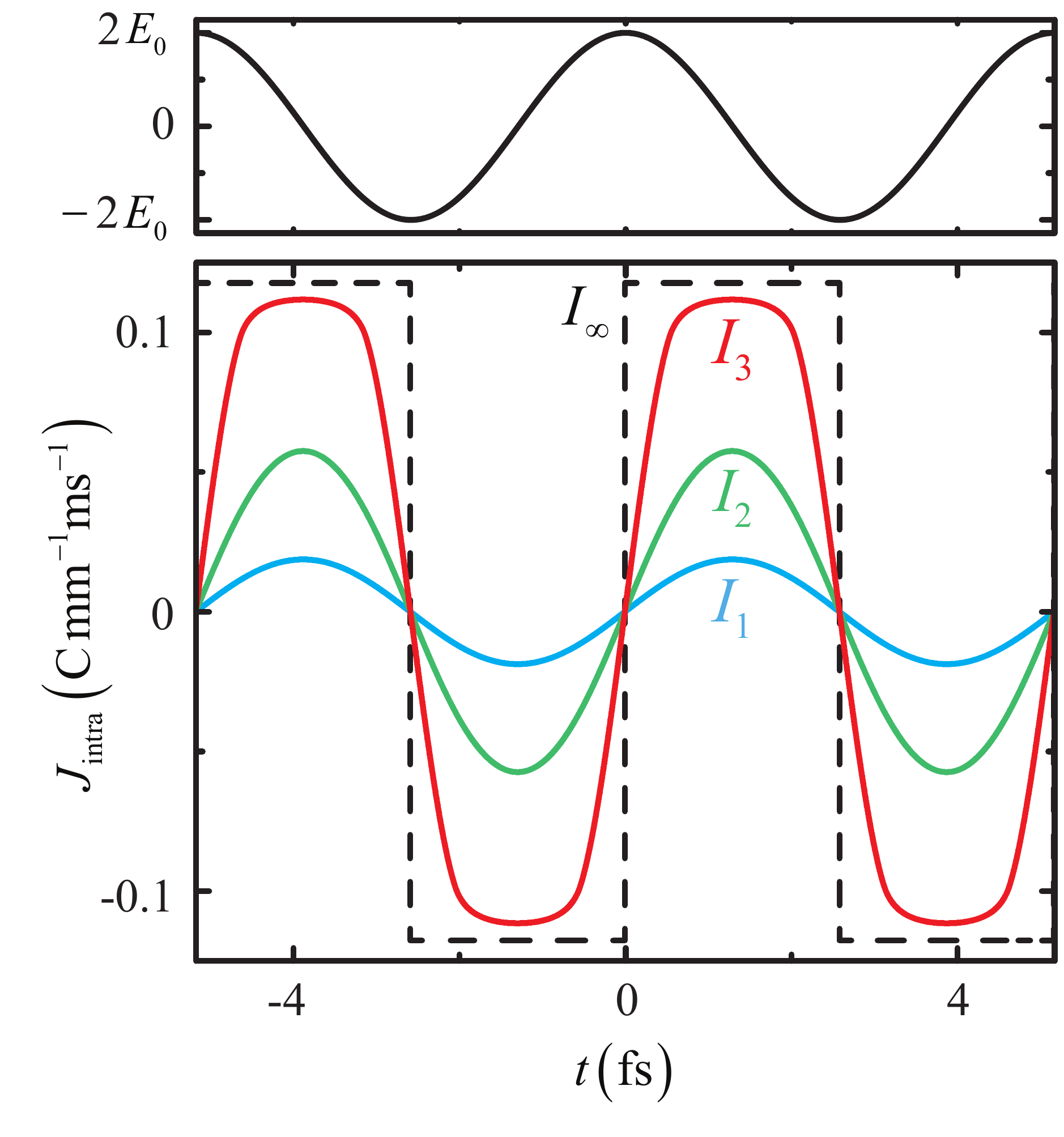}
\caption{(Color online) Intraband surface current density $J_{\rm intra}(t)$ (lower panel) induced in extended graphene by a harmonic external electric field (upper panel) with maximum amplitude $2 E_0 = \sqrt{2I_0/\epsilon_0 c}$ and optical frequency $\omega=2\pi c/\lambda$, where $\lambda=1550\,$nm is the optical excitation wavelength. In the lower panel we provide results for incident intensities $I_0 =10$, 100, and 1000\,GW$/$cm$^2$, compared with the maximum achievable current density $J_{\rm max}(t)$ in the $I_0\rightarrow\infty$ limit (dashed curve). We assume a Fermi energy $E_{\rm F}=1$\,eV and zero temperature.}
\label{Fig2}
\end{center}
\end{figure}

\begin{figure*}[t]
\centering
\begin{center}
\includegraphics[width=\textwidth]{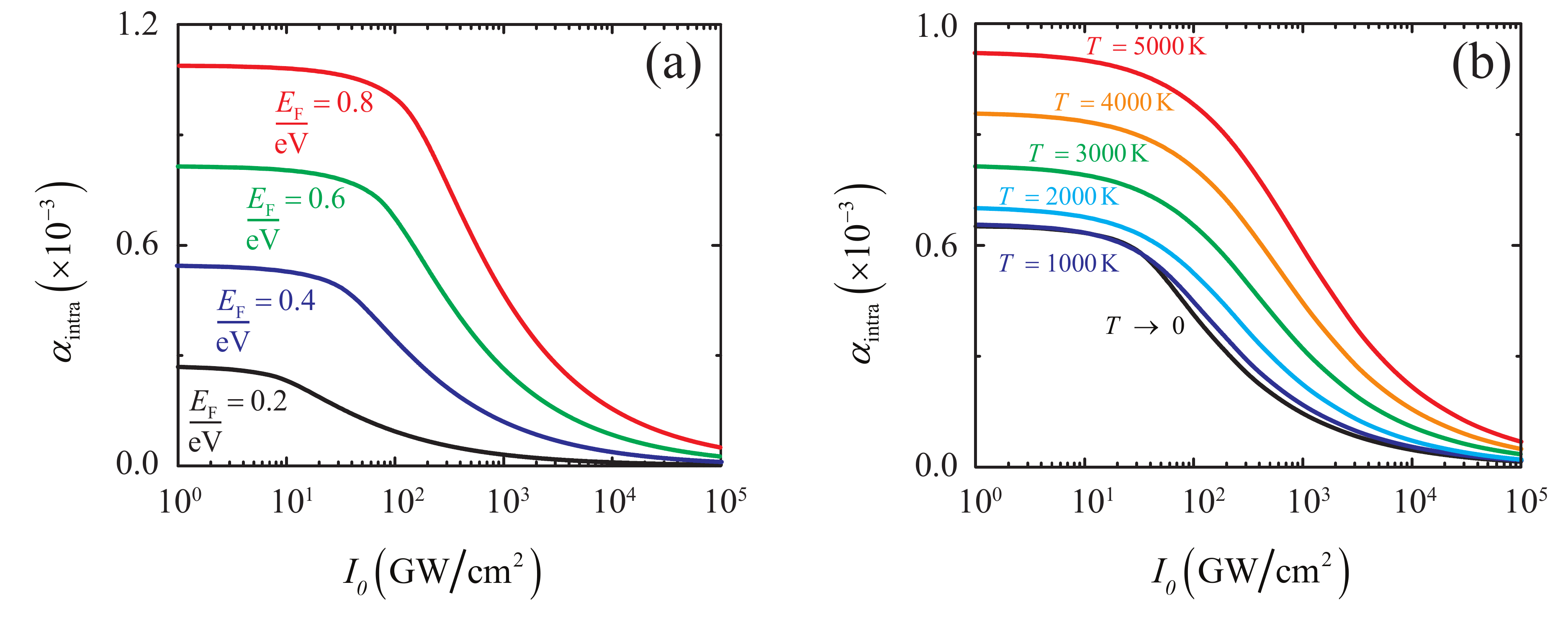}
\caption{(Color online) Intraband absorption coefficient $\alpha_{\rm intra}$ as a function of incident light intensity $I_0$ for {\bf (a)} several Fermi energies $E_{\rm F}$ at zero temperature and {\bf (b)} several electron temperatures $T$ and fixed chemical potential $\mu=0.4\,$eV. The light wavelength is $1550\,$nm.}
\label{Fig3}
\end{center}
\end{figure*}

\noindent This modified expression effectively accounts for intraband electron collisions, and in particular, the direction angle of the electron momentum now becomes $\theta_{\bf p}(t) = \arctan\{p_y/[p_x + e a(t)]\}$. Incidentally, the $\tau\rightarrow\infty$ limit of $\mbox{\boldmath${\pi}$}(t)$ readily reduces to the expression ${\bf p} + e {\bf A}(t)$ that was used in Sec.\ \ref{theoreticalmodel}.

In the limit of vanishing temperature ($T\rightarrow 0$) the chemical potential coincides with the Fermi energy $\mu = E_{\rm F}$, the upper band occupation distribution becomes 
${\cal N}({\bf p})\rightarrow - \Theta(p_{\rm F} - p)$, where $\Theta$ is the Heaviside step function, and the integral in Eq.\ (\ref{IntraIntegralEq}) can be solved analytically [see Appendix\ \ref{A1}]:
\begin{align}
&{\bf J}_{\rm intra}(t)= {\bf \hat{x}} \frac{-2eE_{\rm F}^2}{\pi^2\hbar^2v_{\rm F}} \sqrt{1+[ea(t)/p_{\rm F}]^2} \label{IntraCurrentT0} \\
& \times\int_0^{\pi/2} \cos\phi\left[\sqrt{1+f(t)\cos\phi}-\sqrt{1-f(t)\cos\phi}\right]d\phi, \nonumber
\end{align}
where $p_{\rm F} = E_{\rm F}/v_{\rm F}$ is the Fermi momentum, $f(t)=2ea(t)p_{\rm F}/\{p_{\rm F}^2+[ea(t)]^2\}$, and the integral over $\phi$ can be expressed in terms of generalized Jacobi elliptic functions \cite{AS1972,DongJPB2013} (see Appendix\ \ref{A1}). We emphasize that the intraband surface current density in Eq.\ (\ref{IntraCurrentT0}) is highly nonlinear when $|E_0| > E_{\rm S}$, where $E_{\rm S} = \omega p_{\rm F}/e$ is the intraband saturation field. Additionally, the series expansion of Eq.\ (\ref{IntraCurrentT0}) in powers of the electric field amplitude $E_0$,
\begin{align}
& {\bf J_{\rm intra}}(t) \simeq \frac{e^2}{\pi\hbar^2}{\rm Re} \left\{ \frac{2iE_{\rm F}E_0{\rm e}^{-i\omega t}}{\omega+i\tau^{-1}} +  \frac{ie^2v_{\rm F}^2E_0^3{\rm e}^{-3i\omega t}}{E_{\rm F}(\omega+i\tau^{-1})^3} \right. \nonumber \\
& \left. - \frac{3ie^2v_{\rm F}^2|E_0|^2E_0{\rm e}^{-i\omega t}}{E_{\rm F}(\omega-i\tau^{-1})(\omega+i\tau^{-1})^2} \right\} {\bf \hat{x}} +  {\cal O} \left(E_0^5\right),
\end{align}
fully reproduces the result obtained by means of the Boltzmann transport equation \cite{Peres2014,Mikhailov2007,Mikhailov2008,Mikhailovv2008}.
In order to illustrate intraband saturable absorption of extended graphene, we plot in Fig.\ \ref{Fig2} the intraband current density calculated from Eq.\ (\ref{IntraCurrentT0}) for several values of the incident light intensity $I_0 = 2\epsilon_0 c |E_0|^2$, where $\epsilon_0$ is the vacuum permittivity. Notice that, when $I_0\sim I_{\rm S}^{\rm intra}$, where $I_{\rm S}^{\rm intra}=(1/2)\epsilon_0 c |E_{\rm S}|^2$  is the saturation intensity (e.g., $I_{\rm S}^{\rm intra}\simeq196\,$GW$/$cm$^2$ at $\lambda=1550\,$nm), the current density also saturates, acquiring a square-like-wave temporal profile. This can be verified analytically upon examination of the $E_0\rightarrow \infty$ limit of Eq.\ (\ref{IntraCurrentT0}), which yields the maximum achievable surface current density in doped graphene,
\begin{equation}
{\bf J}_{\rm max}(t) = - e N v_{\rm F} ~ {\rm sign} [a(t)]{\bf \hat{x}}, 
\end{equation}
where $N=p_{\rm F}^2/\pi \hbar^2$ is the density of doping electrons. Then, as a consequence of current saturation, intraband absorption also saturates. We describe this effect quantitatively by defining an intraband absorption coefficient as the ratio of the time average of the absorbed power, which is simply evaluated over a single optical cycle, to the incident-light intensity,
\begin{equation}
\alpha_{\rm intra} \equiv \frac{ \int_{-\pi/\omega}^{+\pi/\omega} {\bf J}_{\rm intra}(t) \cdot {\bf E}(t) dt }{ ( 2\pi / \omega ) I_0}. \label{IntraAbsCoeffEq}
\end{equation}
This quantity reaches its maximum value $\alpha_{\rm max} = 4E_{\rm F}/[137\hbar\tau(\omega^2+\tau^{-2})]$ at low intensities and zero temperature. In contrast, it vanishes as $\alpha_{\rm intra}\simeq 1/\sqrt{I_0}$ in the limit of large incident intensity $I_0$. We further examine the $I_0$ dependence of $\alpha_{\rm intra}(I_0)$ for several Fermi energies [Fig.\ \ref{Fig3}(a)] and electron temperatures [Fig.\ \ref{Fig3}(b)] by numerically solving Eqs.\ (\ref{IntraIntegralEq}) and (\ref{IntraAbsCoeffEq}). A strong dependence on Fermi energy is observed, as well as large thermal modulation at high electron temperatures exceeding $T\sim5000$\,K. We find that the depth of thermal modulation in the intraband absorption is roughly proportional to the Fermi energy and vanishes in undoped graphene. Typical intensities at which the intraband absorption saturates are of the order $100-1000$\,GW$/$cm$^2$, depending on the doping level. Note that the intraband saturation intensity $I_{\rm S}^{\rm intra}$ scales as the inverse square of the optical wavelength ($I_{\rm S}^{\rm intra} \propto \lambda^{-2}$), thus changing by few orders of magnitude within the optical and near-infrared spectrum.

\section{Interband saturable absorption}

We now turn our attention to the effect of interband transitions on saturable absorption, neglecting intraband dynamics. This contribution becomes dominant for photon energies $\hbar \omega > 2 E_{\rm F}$ (see below). Interestingly, while the optical momentum $e{\bf A}(t)$ was found in Sec.\ \ref{intrabandsat} to produce sizable corrections to the intraband dynamics, it does not affect interband dynamics significantly near its resonant contribution at $p = \hbar\omega / (2 v_{\rm F})$, where the optical momentum is negligible [i.e., $p\gg e A(t)$]. We consequently neglect $A(t)$, so that Eqs.\ (\ref{GBEqs}) reduce to
\begin{subequations}
\begin{align}
\dot{\Gamma}_{\bf p} & =  - \left(\frac{1}{\tau} + 2i\omega_0\right)\Gamma_{\bf p} - \frac{ie}{p} {\rm Re} \left\{E_0 {\rm e}^{-i\omega t} \right\} \sin \phi \;n_{\bf p}, \label{BEq1}\\
\dot{n}_{\bf p}      & = - \frac{1}{\tau} \left[n_{\bf p}-{\cal N}\right] + \frac{4e}{p} {\rm Re} \left\{ E_0 {\rm e}^{-i\omega t} \right\} \sin \phi \;{\rm Im}\{\Gamma_{\bf p}\}, \label{BEq2}
\end{align}
\label{BEqs}
\end{subequations}
where $\Gamma_{\bf p}(t) = \rho_{\bf p}(t) {\rm e}^{-2i\omega_0t}$, $\omega_0 = v_{\rm F} p / \hbar$, and we have introduced a phenomenological relaxation time $\tau$ that encompasses the effect of numerous ultrafast decay channels for out-of-equilibrium electrons into hot carriers and phonons \cite{Johannsen2013,Gierz2013,Brida2013}. We assume for simplicity a single effective relaxation time $\tau$, even though polarization dephasing and electron-hole recombination can take place over different time scales and their actual temporal dependences remain uncertain. Additionally, we use the same symbol $\tau$ for this quantity as in the intraband contribution (Sec.\ \ref{intrabandsat}), although we remark that the relative importance of different channels can differ substantially in both cases. Interestingly, Eqs.\ (\ref{BEqs}) coincide with the traditional Bloch equations for graphene and other two-band systems \cite{KochBook,Malic2011}, which are routinely used to describe saturable absorption and other two-band effects, such as for example self-induced transparency \cite{McCallPhysRev1969,GiessenPRL1998,MariniPRL2013}. To describe interband saturable absorption, we adopt the steady-state Ansatz 
\begin{subequations}
\begin{align}
\Gamma_{\bf p}(t) & = \Gamma_{\bf p}^+ {\rm e}^{i\omega t} + \Gamma_{\bf p}^- {\rm e}^{-i\omega t}, \nonumber\\
n_{\bf p}(t)      & = n_{\bf p}^{(0)} + {\rm Re} \left\{ n_{\bf p}^{(2)} {\rm e}^{-2i\omega t} \right\}. \nonumber
\end{align}
\end{subequations}
Using these expressions and neglecting third-harmonic terms, Eqs.\ (\ref{BEqs}) lead to
\begin{subequations}
\begin{align}
& n_{\bf p}^{(0)} = {\cal N} + 4\xi  \;{\rm Im} \left\{ \frac{1-i\omega\tau}{1-i\omega_+\tau} \Gamma_{\bf p}^-\right\}, \\
& n_{\bf p}^{(2)} = \frac{-4 i\xi(1-i\omega\tau)\Gamma_{\bf p}^-}{(1 - 2i\omega\tau)(1-i\omega_+\tau)}, \\
& \Gamma_{\bf p}^+ = - \frac{1+i\omega_-\tau}{1+i\omega_+\tau } {\Gamma_{\bf p}^-}^*, \\
& \Gamma_{\bf p}^- =\frac{(-i\xi/2)}{1-i\omega_-\tau} \left( n_{\bf p}^{(0)} + \frac{1}{2} n_{\bf p}^{(2)} \right), \label{GMMEQ} 
\end{align}
\label{GMMs}
\end{subequations}
where $\xi=(e\tau E_0/p)\sin\phi$ and $\omega_\pm=\omega\pm2\omega_0$. While an analytical solution to Eqs.\ (\ref{GMMs}) is readily obtained, we omit the resulting expressions, which are rather involved and do not provide much physical insight. The interband current density ${\bf J}_{\rm inter}(t)$ is then obtained from this solution and Eq.\ (\ref{CurDensEq}) through the expression
\begin{align}
{\bf J}_{\rm inter}(t) &= - \frac{2 e v_{\rm F}}{\pi^2\hbar^2} \times  \label{InterBandCurrentEq}\\
                        &\times {\rm Re} \left\{ i {\rm e}^{-i\omega t} \int \sin \phi \left[ \Gamma_{\bf p}^- - \Gamma_{\bf p}^{+*} \right] d^2{\bf p} \right\} {\bf \hat{x}}. \nonumber   
\end{align}

\begin{figure}[t]
\centering
\begin{center}
\includegraphics[width=0.45\textwidth]{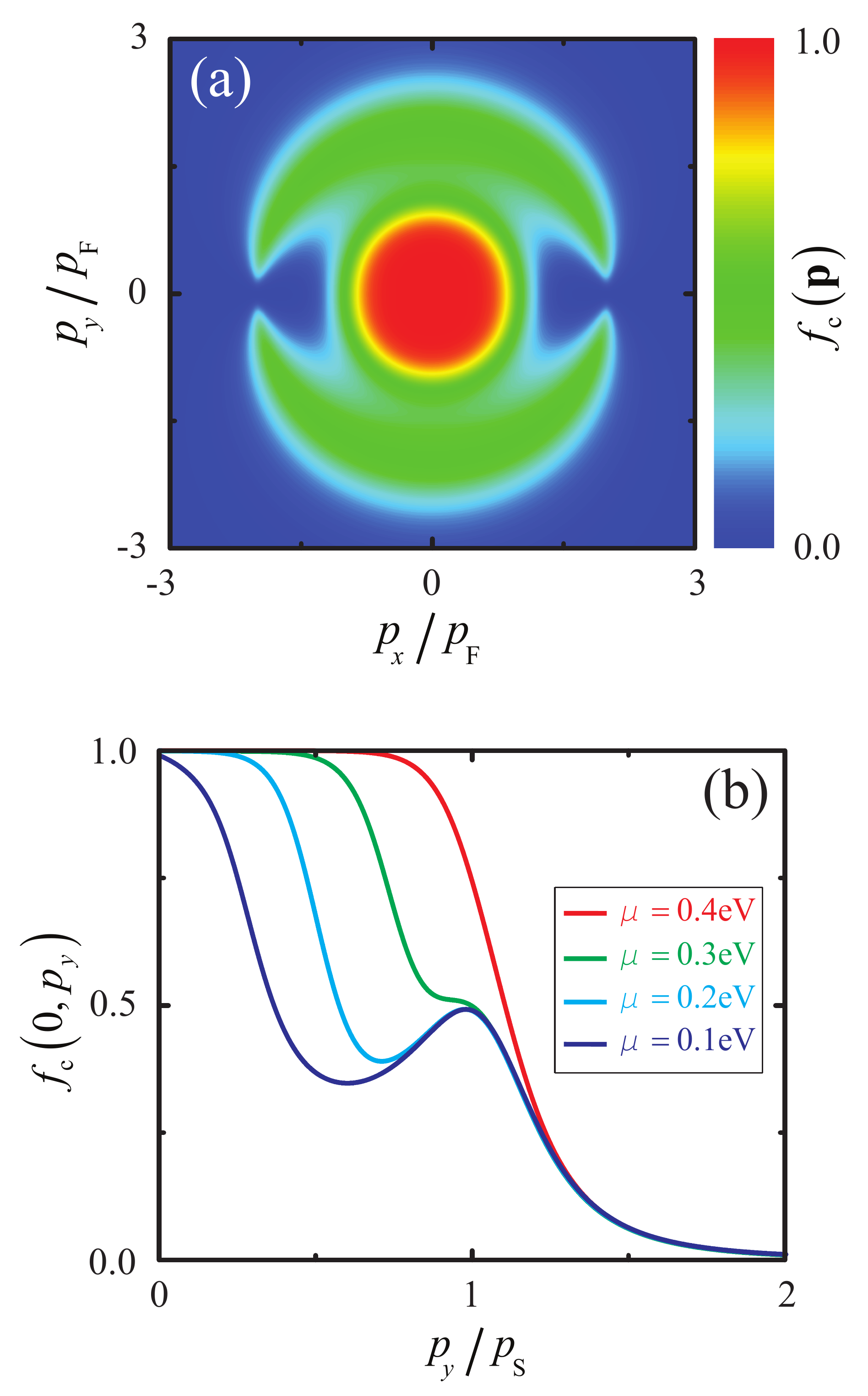}
\caption{(Color online) {\bf (a)} Out-of-equilibrium carrier occupation distribution $f_{\rm c}({\bf p})$ in 2D ${\bf p}$ space at $T=300\,$K for a chemical potential $\mu=0.2\,$eV. The electron momentum components $p_x$ and $p_y$ are normalized to $p_{\rm F} = \mu/v_{\rm F}$. {\bf (b)} Cut along the $p_y$ axis of (a) for several chemical potentials $\mu=0.1-0.4\,$eV, with $p_y$ scaled to $p_{\rm S}=\hbar\omega/2v_{\rm F}$. The light intensity and wavelength are $I_0 = 10\,$GW$/$cm$^2$ and $\lambda = 1550\,$nm in both plots. We assume a phenomenological relaxation time $\tau=22\,$fs.}
\label{Fig4}
\end{center}
\end{figure}

\noindent We finally solve the 2D integral in ${\bf p}$ space numerically. Because we neglect third-harmonic terms, $\Gamma_{\bf p}$ oscillates in time around zero with the same angular frequency $\omega$ as the external field, while $n_{\bf p}$ oscillates with angular frequency $2\omega$ around the steady-state out-of-equilibrium population difference $n_{\bf p}^{(0)}$. Consequently, our approach enables the explicit calculation of the out-of-equilibrium occupation distribution of optically induced free-carriers $f_{\rm c}({\bf p}) = [{\cal F}(p)+{\cal F}(-p)+n_{\bf p}^{(0)}]/2$ [see Figs. \ref{Fig4}(a,b), where we plot $f_{\rm c}({\bf p})$ for several chemical potentials $\mu$ at fixed incident light intensity $I_0 = 10$ GW$/$cm$^2$ and wavelength $\lambda=1550\,$nm].

\begin{figure}[t]
\centering
\begin{center}
\includegraphics[width=0.45\textwidth]{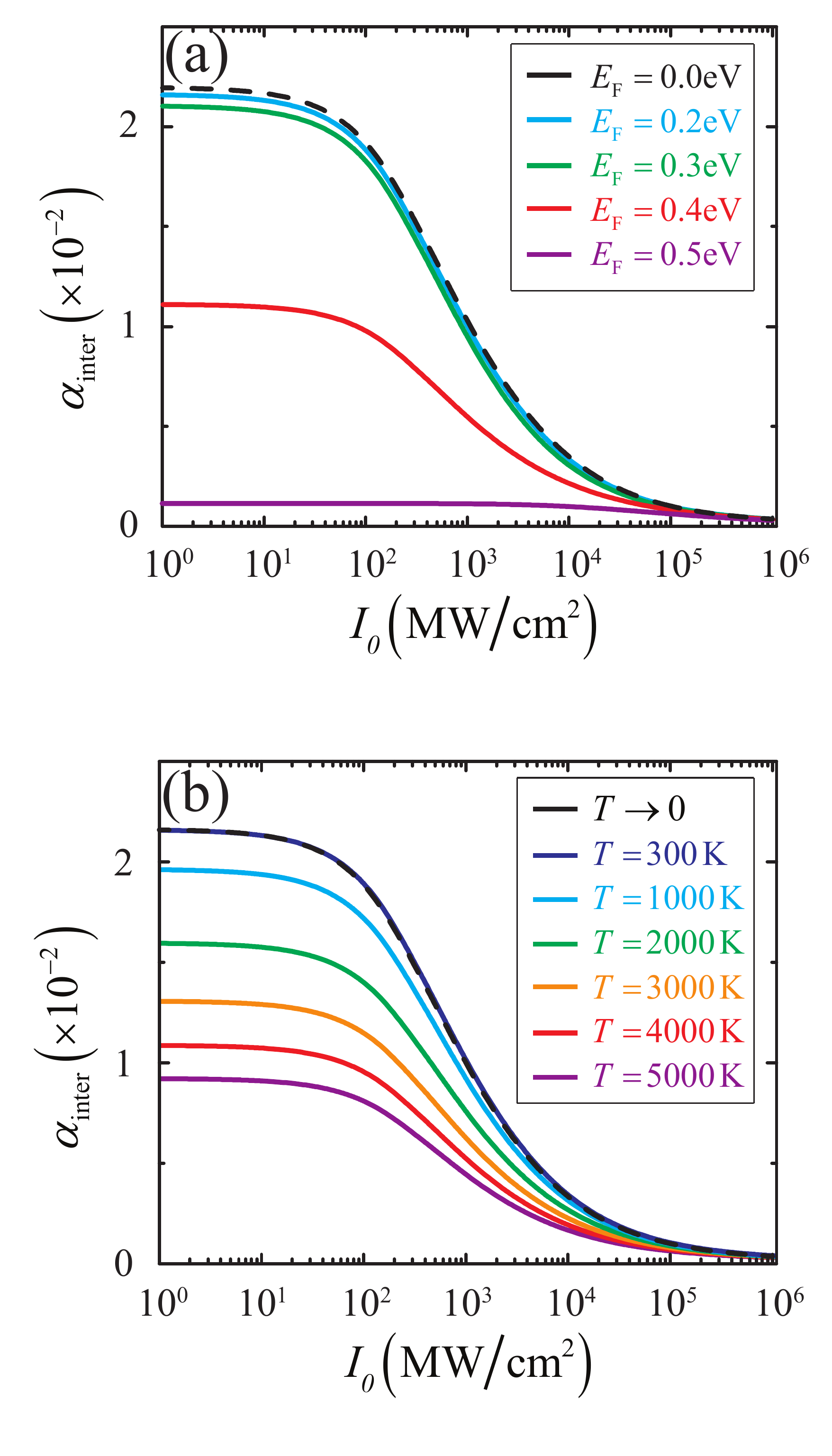}
\caption{(Color online) Interband absorption coefficient $\alpha_{\rm inter}$ as a function of incident light intensity $I_0$ for a wavelength of $1550\,$nm for {\bf (a)} several Fermi energies $E_{\rm F}$ at zero temperature and {\bf (b)} several electron temperatures $T$ and fixed chemical potential $\mu = 0.2$ eV. The inelastic relaxation time is set to $\tau=22\,$fs in all plots.}
\label{Fig5}
\end{center}
\end{figure}

The ${\bf p}$-space steady-state carrier occupation $f_{\rm c}({\bf p})$ is characterized by two lobes along the ${\bf \hat{y}}$ axis, peaked at $(p_x,p_y)=(0, \pm\hbar\omega/2v_{\rm F})$ and surrounding the thermal Fermi-Dirac distribution of electrically doped electrons [see Fig.\ \ref{Fig4}(a)].  When $\hbar\omega<2\mu/\hbar$, the distribution $f_{\rm c}({\bf p})$ stays thermalized, as interband absorption is inhibited by the Pauli exclusion principle. In particular, after irradiation by an ultrashort optical pulse, $f_{\rm c}({\bf p})$ quickly evolves within a time of $\sim\tau=22\,$fs to a different thermal distribution corresponding to an increased electron temperature \cite{Malic2011}. Eventually, it relaxes to the original unperturbed distribution ${\cal F} (p)$ within a time scale $\sim1\,$ps due to electron-phonon scattering \cite{Johannsen2013,Gierz2013,Brida2013}.

\begin{figure}[t]
\centering
\begin{center}
\includegraphics[width=0.47\textwidth]{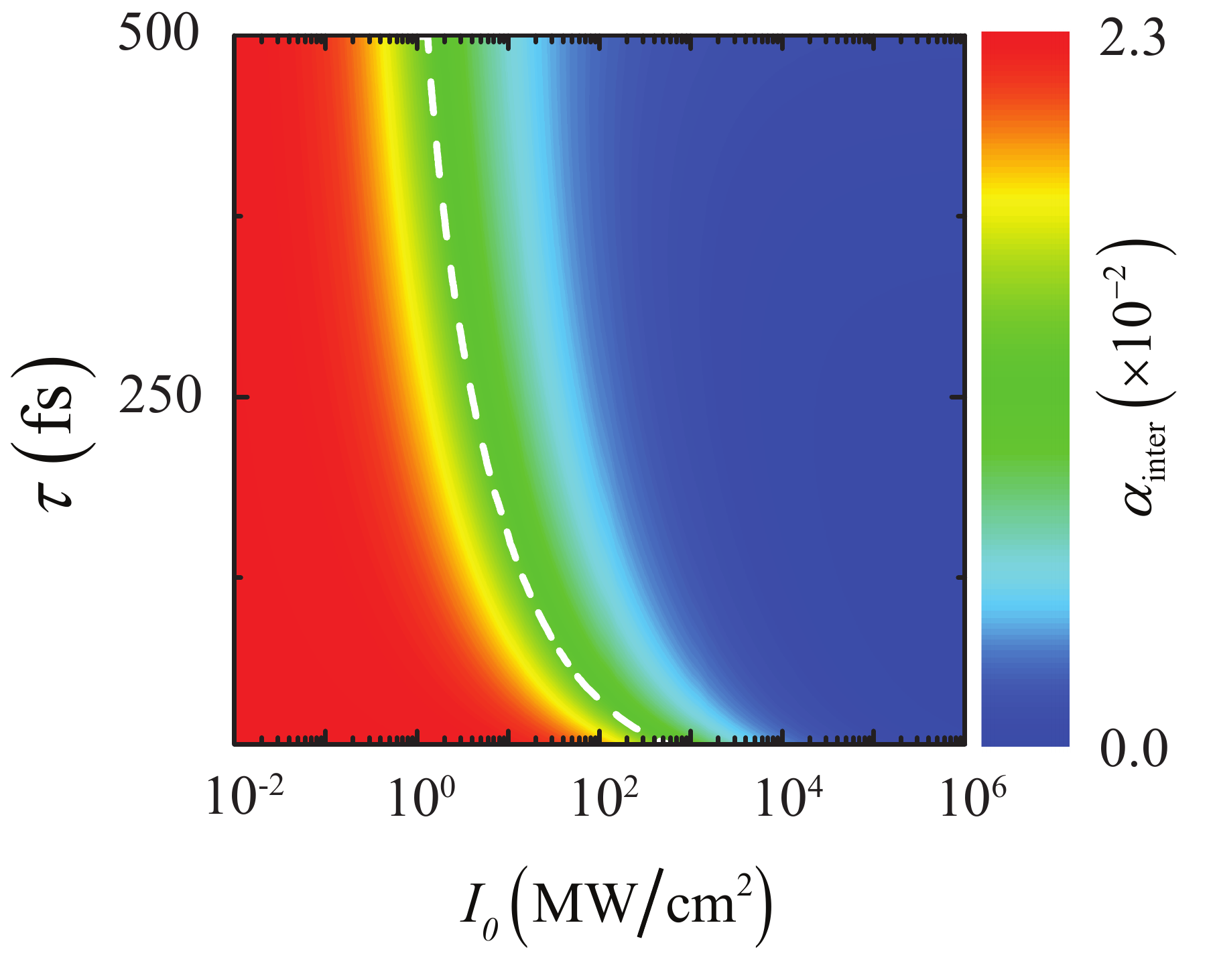}
\caption{(Color online) Dependence of the interband absorption coefficient $\alpha_{\rm inter}$ of undoped graphene on incident light intensity $I_0$ and inelastic relaxation time $\tau$ at zero temperature. The white dashed curve represents the saturation intensity $I_{\rm S}^{\rm inter}(\tau)$ for which $\alpha_{\rm inter}(I_{\rm S}^{\rm inter})=\alpha_{\rm inter}(0) / 2$.}
\label{Fig6}
\end{center}
\end{figure}

As a consequence of the optically induced out-of-equilibrium occupation $f_{\rm c}({\bf p})$ in the upper band, interband absorption also saturates. In simple terms, at high intensities the electron-hole recombinations produced by inelastic collisions balance the light-driven interband transitions, thus leading to absorption quenching. Similar to Sec.\ \ref{intrabandsat}, we attempt to quantitatively describe this effect by defining an interband absorption coefficient as the ratio of the time-averaged absorbed power over an optical cycle to the impinging intensity,
\begin{equation}
\alpha_{\rm inter} \equiv \frac{ \int_{-\pi/\omega}^{+\pi/\omega} {\bf J}_{\rm inter} (t) \cdot {\bf E} (t) dt }{(2\pi/\omega) I_0}. \label{InterAbsCoeffEq}
\end{equation}
We calculate the intensity-dependent interband absorption coefficient $\alpha_{\rm inter}(I_0)$ for several Fermi energies [Fig.\ \ref{Fig5}(a)] and electron temperatures [Fig.\ \ref{Fig5}(b)] by numerically solving Eqs.\ (\ref{InterBandCurrentEq}) and (\ref{InterAbsCoeffEq}). For low intensities and zero temperature, we reproduce the universal absorption law of undoped graphene, $\alpha_{\rm inter}\approx\pi\alpha$ \cite{NBG08,MSW08}, where $\alpha\approx1/137$ is the fine-structure constant. This result reflects a dispersionless linear conductivity $\sigma_0=e^2/4\hbar$. In contrast, the interband absorption vanishes as $\alpha_{\rm inter}\simeq 1/\sqrt{I_0}$ in the limit of large incident intensities $I_0$. This high-intensity behavior is similar to intraband absorption (see Sec.\ \ref{intrabandsat}). We also observe a strong dependence of $\alpha_{\rm inter}(I_0)$ on the Fermi energy, as well as a large thermal modulation at high electron temperatures $T\simeq 2000\,$K. It should be noted that the modulation depth is much higher for interband than for intraband absorption, and additionally, it can be efficiently controlled by changing the Fermi energy. 

\begin{figure}[t]
\centering
\begin{center}
\includegraphics[width=0.47\textwidth]{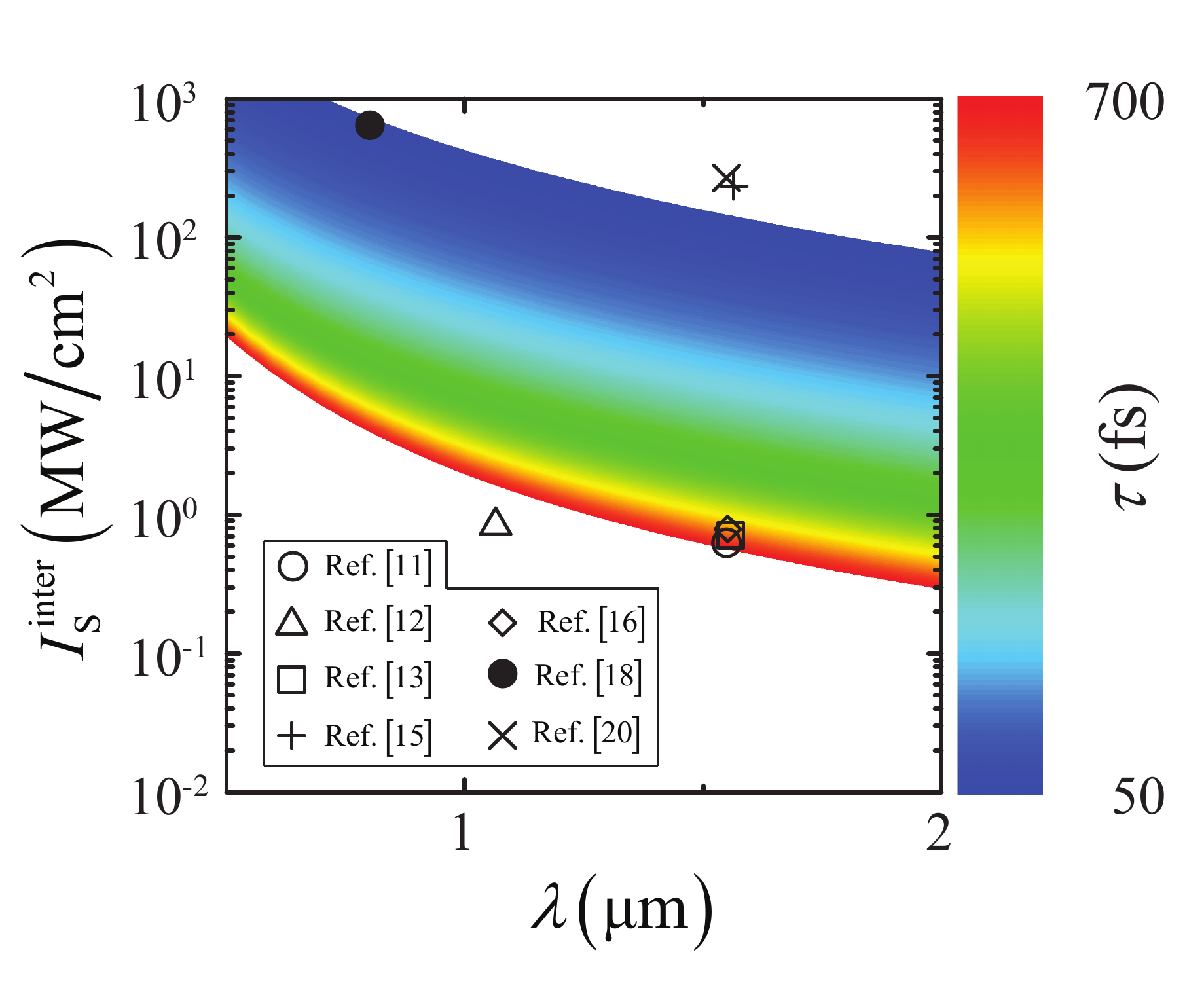}
\caption{(Color online) Predicted interband saturation intensity $I_{\rm S}^{\rm inter}$ (left scale) as a function of optical wavelength $\lambda$ (horizontal axis) and relaxation time $\tau$ (color scale) at zero temperature. Available experimental results for the saturation intensity are indicated by symbols \cite{Hasan2009,Bao2009,Tan2010,Zhangg2010,Xing2010,Martinez2011,Baoo2011,Winzer2012,Baek2013} (see legend).}
\label{Fig7}
\end{center}
\end{figure}

The light intensity required to achieve saturated absorption can be affected by the actual value of the inelastic collision time. For large $\tau$, optically pumped electrons are expected to produce Pauli blocking during a longer time, therefore reducing the interband saturation intensity $I_{\rm S}^{\rm inter}$. A plot of $\alpha_{\rm inter}$ as a function of $I_0$ and $\tau$ [Fig.\ \ref{Fig6}(a), calculated for $\lambda=1550\,$nm] confirms this intuition and further reveals that $I_{\rm S}^{\rm inter}$ varies over a wide range ($1-100\,$MW$/$cm$^2$) when $\tau$ evolves within a range compatible with reported measurements of the electron inelastic lifetime in graphene samples of different qualities. But more importantly, the actual values of $I_{\rm S}^{\rm inter}$ are much smaller than the characteristic intraband saturation intensities $I_{\rm S}^{\rm intra}$ derived in Sec.\ \ref{intrabandsat}. The interband saturation intensity $I_{\rm S}^{\rm inter}$ heavily depends on the optical wavelength $\lambda$ [see Fig.\ \ref{Fig7}, where we plot $I_{\rm S}^{\rm inter}(\lambda)$ for several relaxation times $\tau$], varying by several orders of magnitude within the optical and near-infrared spectrum. 

\begin{figure}[t]
\centering
\begin{center}
\includegraphics[width=0.45\textwidth]{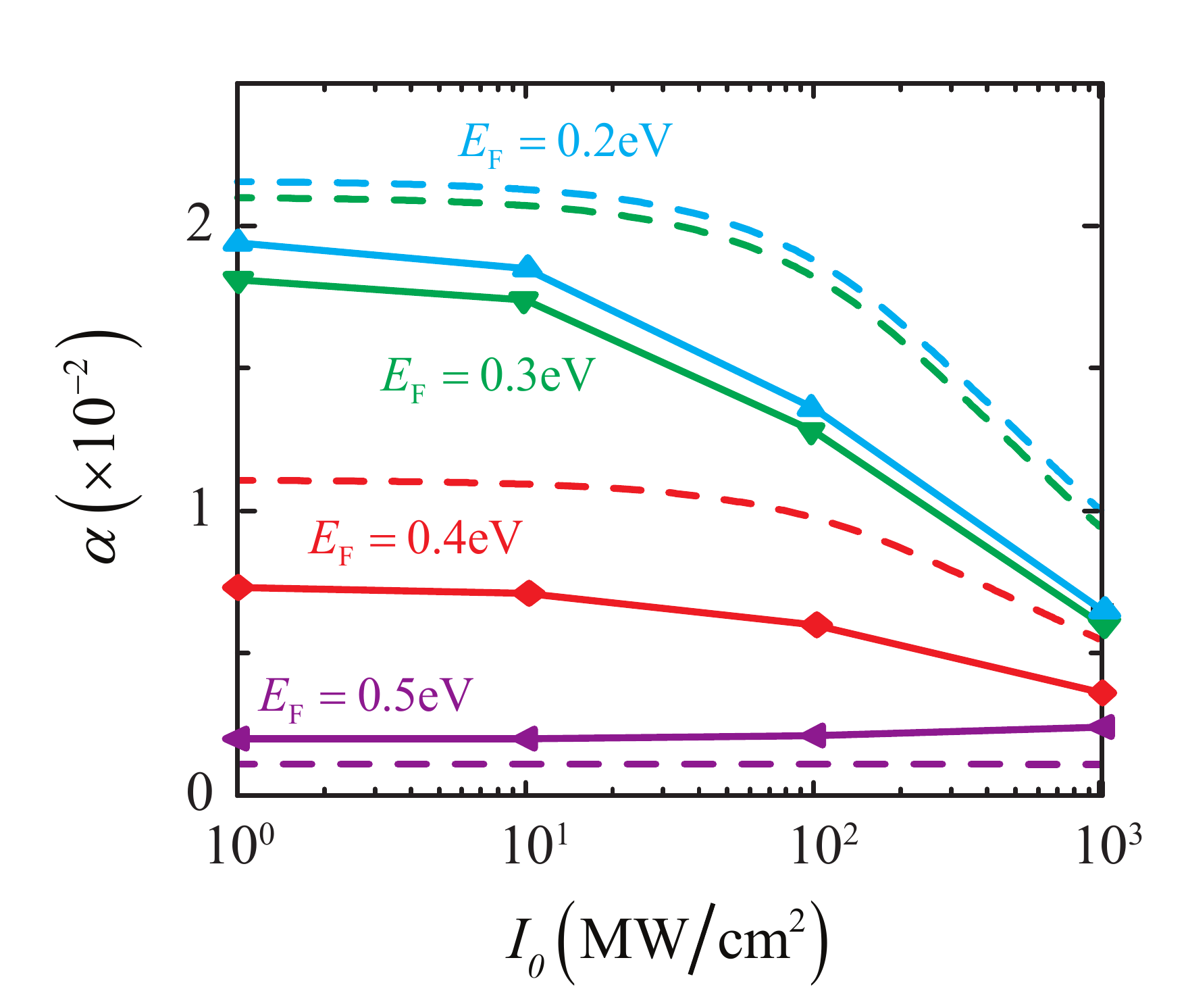}
\caption{(Color online) Comparison between numerical atomistic simulations (full $\alpha$, solid curves and symbols) and the analytical MDF model for the interband absorption coefficient ($\alpha_{\rm inter}$, dashed curves) as a function of incident light intensity $I_0$ for a wavelength of $1550\,$nm and several Fermi energies $E_{\rm F}$ at zero temperature.}
\label{Fig8}
\end{center}
\end{figure}

\section{Finite-size and atomistic effects}

Although we have provided a comprehensive theory of saturable absorption in extended graphene, practical devices operating on this principle must be finite in size. This modifies the electronic structure of the material and thus its optical response. While this effect is only expected to play a significant role in structures with dimensions less than a few tens of nanometers \cite{paper183}, we contrast next the above theoretical description for extended graphene with results obtained from an atomistic approach for finite structures. Specifically, following the methods of Refs.\,\cite{Cox2014,paper269}, we simulate the intensity-dependent optical response of one-dimensional graphene nanoribbons by numerically solving the single-particle density matrix equation of motion in the time domain, using a tight-binding Hamiltonian for the $\pi$-band electronic structure along with a self-consistent electron-electron (Hartree) interaction potential (see Appendix \ref{A2} for further details).

Remarkably, for a nanoribbon of only $\sim20$\,nm width, we find the atomistic simulations of the intensity-dependent absorbed power under cw illumination to be in excellent quantitative agreement with that for extended graphene based on the MDF picture (see Fig.\ \ref{Fig8}, where we compare results from the two methods for several Fermi energies at an excitation wavelength of $1550\,$nm). Despite this agreement on the saturation intensity, the unsaturated absorption coefficients in the weak intensity limit differ slightly, presumably due to finite-size effects that are captured in the atomistic approach but inherently absent in the MDF picture.

\begin{figure}[t]
\centering
\begin{center}
\includegraphics[width = 0.45\textwidth]{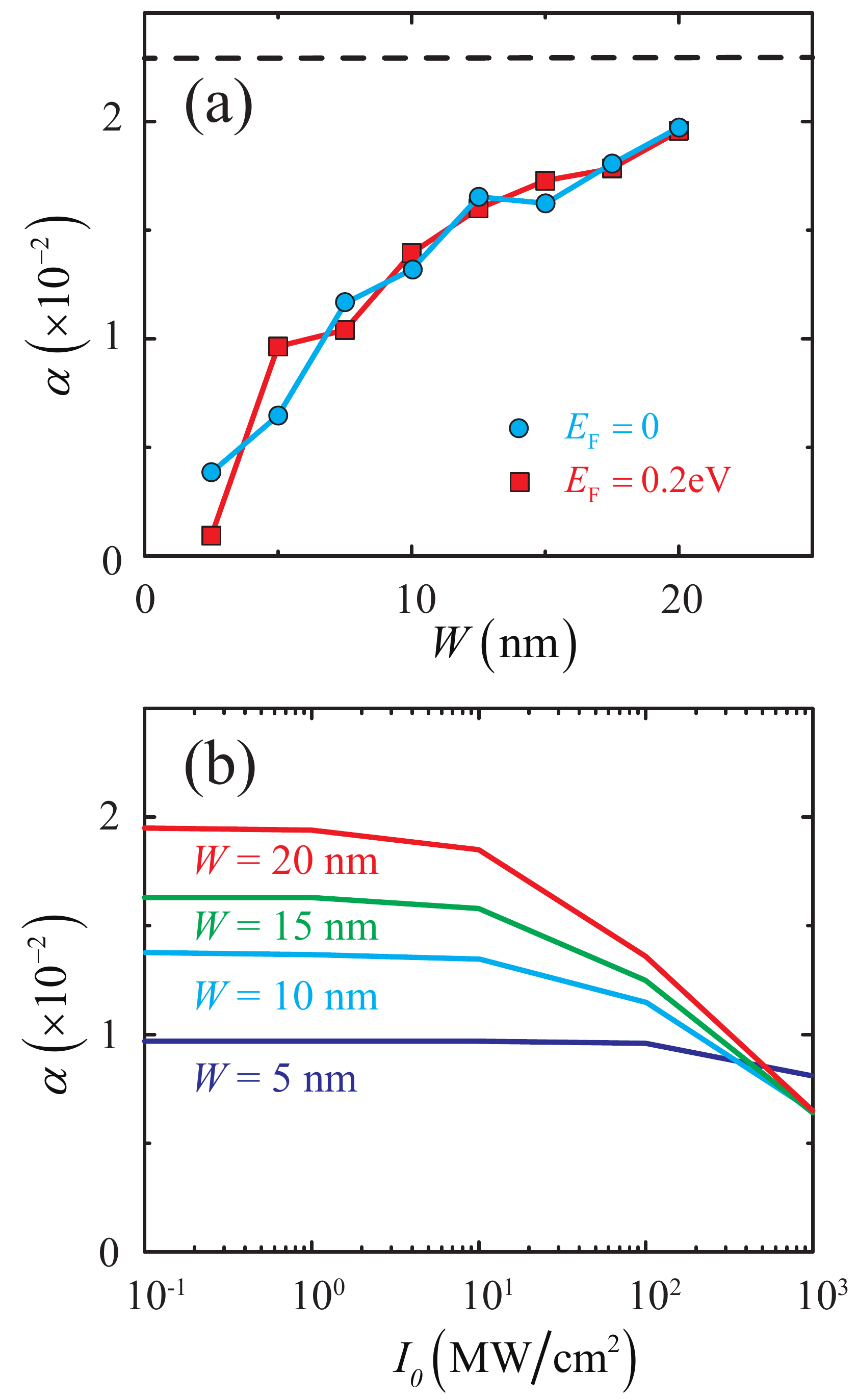}
\caption{(Color online) Results from atomistic simulations of graphene ribbons depicting the dependence of the absorption coefficient $\alpha$ either ({\bf a}) on ribbon width $W$ for two values of $E_{\rm F}$ and ({\bf b}) on impinging light intensity $I_0$ for a fixed Fermi energy $E_{\rm F} = 0.2$ eV and several values of $W$. The dashed line in ({\bf a}) indicates the universal absorption limit $\alpha \approx \pi/137$.}
\label{Fig9}
\end{center}
\end{figure}

Interestingly, for a Fermi energy of $0.5\,$eV, the atomistic approach predicts an intensity-dependent increase in absorption rather than saturation. This discrepancy could originate in two-photon absorption processes: although one-photon absorption (for which the analytical derivation of the previous section is valid) is quenched at $\lambda = 1.55\,\mu$m and $E_{\rm F}=0.5\,$eV, two-photon absorption leads to an increase in absorption at very high intensities. A detailed description of two-photon absorption goes however beyond the scope of the present discussion. 

Figure \ref{Fig9} depicts the results of atomistic simulations for the dependence of the absorption coefficient $\alpha$ on the ribbon width $W$ and the impinging light intensity $I_0$, which are shown to converge for widths of a few tens of nanometers. This corroborates that finite-size effects can be indeed neglected for larger graphene structures (i.e., flakes of 10's nm in size).

\begin{figure}[t]
\centering
\begin{center}
\includegraphics[width=0.45\textwidth]{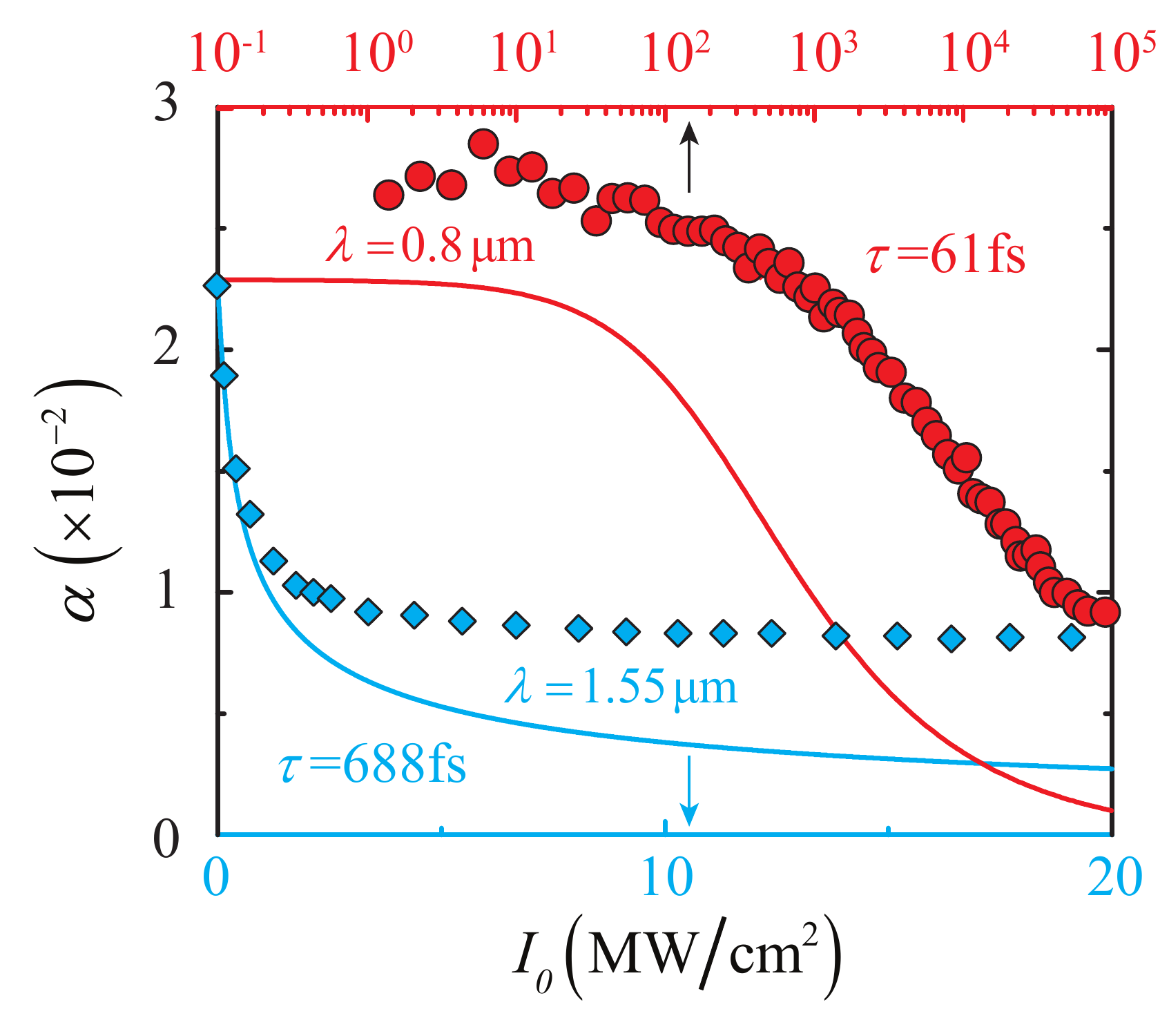}
\caption{(Color online) Comparison of theory (curves) and experiments (symbols) for the intensity-dependent absorption coefficient $\alpha_{\rm inter}(I_0)$ at two different optical wavelengths: $\lambda = 800\,$nm (red curves and symbols) and $\lambda = 1550\,$nm (blue curves and symbols). Experimental results are taken from Refs.\ \cite{Baoo2011} (rombic markers) and \cite{Baek2013} (circle markers).}
\label{Fig10}
\end{center}
\end{figure}

\section{Comparison with experiments}

We are now ready to compare our theoretical results with experimental data available in the literature. Most experimental studies exploit graphene as a saturable absorber for PML applications, and thus focus on undoped or poorly doped samples, for which only interband absorption becomes relevant. Surprisingly, we find enormous variations in reported measurements, even for experiments conducted with the same light wavelength, so we attribute this dispersion in observed results to the different qualities of the graphene samples and experimental conditions. This intuition is supported by the calculations presented in Fig.\ \ref{Fig7} (see above), in which the saturation intensity for SA is shown to be highly dependent on the intrinsic relaxation time $\tau$, which is directly affected by the graphene sample quality. In contrast, the saturation intensity is less sensitive to doping (for $E_{\rm F}<2\hbar\omega$), although these parameters have a considerable effect on the modulation depth.

In Fig.\ \ref{Fig7} we compare the calculated saturation intensity (color plot) to experimental measurements (symbols) for several optical wavelengths and relaxation times, while in Fig.\ \ref{Fig10} we contrast the predicted intensity-dependent absorption coefficient (curves) with experimental data (symbols). We find good agreement with several experimental results \cite{Bao2009,Zhangg2010,Baoo2011,Baek2013} by assuming reasonable values for the relaxation times, ranging from $50\,$fs to $700\,$fs, while other experiments \cite{Tan2010,Xing2010,Martinez2011,Winzer2012,Sun2010} can only be reproduced either with extremely low relaxation times $<50\,$fs, presumably corresponding to low-quality samples, or with extremely high relaxation times $>1\,$ps. We also note that a non-vanishing background absorption, which does not saturate within the considered range of illumination intensities, is usually present in the aforementioned experimental results [see Fig.\ \ref{Fig10}]. This effect appears to be dependent on the number of graphene layers \cite{Bao2009} and is found to vanish for pristine single-layer extended graphene \cite{Baoo2011}.

\section{Concluding remarks}

In summary, we have developed a non-perturbative description of both intraband and interband contributions to saturable absorption in extended graphene by modeling conduction electrons as 2D massless Dirac fermions coupled to an external optical field. The Dirac equation describing the time evolution of these electrons is recast in the form of generalized Bloch equations, which we solve analytically. Remarkably, the interband contribution presents saturation at unusually low light intensities when compared with other materials, or with the intraband saturation. Additionally, we find a significant dependence of these effects on Fermi energy, providing an active mechanism of control over saturable absorption in the carbon layer via electrical doping modulation. In contrast, the electron temperature $T$ does not play a relevant role, unless $k_{\rm B}T$ exceeds the Fermi energy or the incident photon energy. Nonetheless, $T$ can reach that regime under intense optical pumping, and therefore, this is an effect that must be considered when illuminating with long, intense light pulses.

We attribute the strong saturable absorption in graphene to its peculiar electronic band structure, and in particular, to the combination of its linear dispersion relation and the vanishing of the density of states at the Dirac point. Consequently, both doping and optical transitions in the presence of strong optical fields produce large modifications in the populations of electronic states, thus resulting in substantial variations in momentum and producing radical changes in the conductivity. We therefore expect a similarly low threshold for saturable absorption in other nanoscale materials that present Dirac cones or other exotic electronic structures characterized by a low density of electronic states at the Fermi level. Band-structure engineering is then a direction to explore in nanographenes, fullerenes, and carbon nanotubes, as well as in other van der Waals atomic-layer materials and its derivatives.

Interband absorption dominates under the conditions explored in currently available experiments, which are in good agreement with our analytical theory. In contrast, intraband saturation absorption is not widely studied in the literature, and we predict it to emerge only at very high light intensities $\sim100-1000$\,GW$/$cm$^2$, depending on the doping level. However, the intraband saturation intensity scales as the inverse square of the optical wavelength, and thus should become important at Terahertz frequencies, particularly if it is enhanced by localized graphene plasmons in nano-islands/ribbons \cite{Jadidi2016}.

Indeed, while we find good agreement with an atomistic description of saturable absorption, including finite-size effects under nonresonant conditions, we anticipate that in general the enhanced near-fields associated with localized plasmons in graphene nanostructures should contribute significantly to reduce the saturation intensity and enlarge the modulation depth of SA. A detailed study of plasmon-enhanced SA is still needed to explore how far down the saturation intensity can be pushed.

Overall, the extraordinarily low intensity threshold for saturable absorption in graphene, combined with its electrical tunability, offers great potential for photonic applications such as mode-locking in graphene-clad fibre lasers and graphene-based random lasers.

\acknowledgments

This work has been partially supported by the Spanish MINECO (MAT2014-59096-P and SEV2015-0522) and the European Commission (Graphene Flagship CNECT-ICT-604391 and FP7-ICT-2013-613024-GRASP).  

\appendix

\section{Derivation of Eq.\ (\ref{IntraCurrentT0})}
\label{A1}

Here, we provide further details on the derivation of the intraband current [Eq.\ (\ref{IntraCurrentT0})]. At zero temperature the chemical potential coincides with the Fermi energy ($\mu = E_{\rm F}$) and ${\cal N}({\bf p})=-\Theta(p_{\rm F} - p)$ (we assume doping with holes). Inserting this and the expression given in the main text for $\cos\theta_{\bf p}$ into Eq.\ (\ref{IntraIntegralEq}), we find
\begin{align}
&{\bf J}_{\rm intra}(t) = - \frac{ev_{\rm F}}{\pi^2\hbar^2} {\bf \hat{x}} \int \Theta(p - p_{\rm F}) \cos \theta_{\bf p}(t) d^2{\bf p} \\
& = - \frac{ev_{\rm F}}{\pi^2\hbar^2} {\bf \hat{x}} \int_{-p_{\rm F}}^{p_{\rm F}} dp_y \int_{-\sqrt{p_{\rm F}^2-p_y^2}}^{\sqrt{p_{\rm F}^2-p_y^2}}dp_x \frac{p_x+ea(t)}{\sqrt{[p_x+ea(t)]^2+p_y^2}}. \nonumber
\end{align}
Now, we perform the $p_x$ integral analytically and define $f(t)=2ea(t)p_{\rm F}/\{p_{\rm F}^2+[ea(t)]^2\}$ to obtain Eq.\ (\ref{IntraCurrentT0}). Finally, the $\phi$ integral in Eq.\ (\ref{IntraCurrentT0}) gives rise to the analytical expression
\begin{align}
{\bf J}_{\rm intra}(t)= {\bf \hat{x}} \frac{-2eE_{\rm F}^2}{\pi^2\hbar^2v_{\rm F}} \sqrt{1+[ea(t)/p_{\rm F}]^2} \;{\cal I}[f(t)] ,
\end{align}
where
\begin{align}
& {\cal I} = \frac{2}{3f} \left\{ \frac{f^2-1}{\sqrt{1+f}} \left[ {\cal F} \left( \frac{\pi}{4}\left|\frac{2f}{f+1}\right. \right) - {\cal F} \left( 0\left|\frac{2f}{f+1}\right. \right) \right] \right. \nonumber \\
& \left. + \frac{f^2-1}{\sqrt{1-f}} \left[ {\cal F} \left( \frac{\pi}{4}\left|\frac{2f}{f-1}\right. \right) - {\cal F} \left( 0\left|\frac{2f}{f-1}\right. \right) \right] \right. \nonumber \\ 
& \left.  + \sqrt{1+f} \left[ {\cal E} \left( \frac{\pi}{4}\left|\frac{2f}{f+1}\right. \right) - {\cal E} \left( 0\left|\frac{2f}{f+1}\right. \right) \right] \right. \nonumber \\ 
& \left.  + \sqrt{1-f} \left[ {\cal E} \left( \frac{\pi}{4}\left|\frac{2f}{f-1}\right. \right) - {\cal E} \left( 0\left|\frac{2f}{f-1}\right. \right) \right]\right\} 
\end{align}
and 
\begin{align}
& {\cal F}(\left.\phi\right|m) = \int_0^\phi \frac{1}{\sqrt{1-m \sin^2\theta}} ~ d\theta, \\
& {\cal E}(\left.\phi\right|m) = \int_0^\phi \sqrt{1-m \sin^2\theta} ~ d\theta
\end{align} 
are the generalized Jacobi elliptic functions of first and second kind \cite{AS1972}. 

\section{Atomistic quantum-mechanical simulations}
\label{A2}

We consider a graphene ribbon containing $N\rightarrow\infty$ unit cells with period $b$ along its direction of translational symmetry. Following a previously reported procedure \cite{paper183, Cox2014, paper269}, we construct the corresponding one-electron wave functions from a tight-binding Hamiltonian $H_\text{TB}$ (assuming a nearest-neighbor hopping energy of 2.8\,eV) as $\ket{j,k}=\sum_{l,m}a_{jl,k}\text{e}^{i kmb}\ket{l,m}/\sqrt{N}$, where $j$ denotes the band index, $k$ is the in-plane Bloch wave vector along the ribbon, $\ket{l,m}$ is the 2p carbon orbital at site $\textbf{R}_l$ in unit cell $m$, and $a_{jl,k}$ are complex expansion coefficients. The optical response is simulated via direct numerical integration of the single-electron density matrix equation of motion,
\begin{equation}\label{rhoeom}
\frac{\partial \rho}{\partial t}=-\frac{i}{\hbar}[H_\text{TB}-e\phi,\rho]-\frac{1}{2\tau}\left(\rho-\rho^0\right),
\end{equation}
where $\phi=-\textbf{R}_l\cdot\textbf{E}(t)-2e\sum_{l',m'}v_{ll',mm'}\rho_{l'l',m'm'}$ is the self-consistent electric potential, with $v_{ll',mm'}$ denoting the Coulomb interaction between atom $l$ in unit cell $m$ and atom $l'$ in unit cell $m'$. Relaxation in Eq.\ (\ref{rhoeom}) brings us back to the equilibrium density matrix $\rho^0$ at a rate $\tau^{-1}$. In the state representation, the relaxed matrix elements $\rho^{0}_{jj',kk'}=f_{j,k}\delta_{jj'}\delta_{kk'}$ are constructed by populating electron states according to the Fermi-Dirac distribution (see occupation numbers $f_{j,k}$). Here we consider light impinging along the graphene plane normal, with polarization $\textbf{E}_0$ directed across the ribbon [i.e., the incident light field is $\textbf{E}(t)={\rm Re}\left\{\textbf{E}_0\exp(-i\omega t)\right\}$]. Upon integration of Eq.\ (\ref{rhoeom}), the time-dependent elements $\rho_{ll}$ yield the induced dipole moment per unit length of the nanoribbon
\begin{equation}
\mathfrak{p}(t)=-2e\sum_l \textbf{R}_l\cdot\textbf{E}_0 \int^{\pi/b}_{-\pi/b}  \frac{dk}{2\pi}\left( \rho_{ll,kk}-\rho^0_{ll,kk} \right).
\end{equation}
The absorption coefficient is then calculated from the optical cross-section normalized to the ribbon area, $\alpha=4\pi\omega \text{Im}\{\mathfrak{p}(\omega)\} / c|E_0|^2 W$,
where $W$ is the nanoribbon width and $\mathfrak{p}(\omega)=(\omega/2\pi)\int^{t_f}_{t_f-2\pi/\omega}dt\,\mathfrak{p}(t)\text{e}^{i\omega t}$ is the Fourier transform of $\mathfrak{p}(t)$ over a single optical cycle of the incident cw field.

\end{document}